\documentclass[12pt]{iopart}
\newcommand{\Lp}{L_x+i\eta L_y}
\newcommand{\sqe}{\sqrt{1-\eta^2}}
\newcommand{\mylims}{\stackrel{\scriptscriptstyle N\gg 1}
{\longrightarrow}}
\usepackage{graphics}
\begin{document}
\title[Geometrical properties of intelligent spin states]
{Geometrical properties of intelligent spin states 
and time evolution of coherent states}
\author{R Arvieu\dag\footnote[4]{E-mail: arvieu@in2p3.fr
and rozmej@tytan.umcs.lublin.pl}  and P Rozmej\ddag\S} 

\address{\dag Institut des Sciences Nucl\'eaires, F-38026 Grenoble, France}
\address{\ddag Instytut Fizyki, Uniwersytet MCS, 20-031 Lublin, Poland}
\address{\S Gesellschaft f\"ur Schwerionenforschung, 
D-64220 Darmstadt, Germany}
\begin{abstract}
 We remind the properties of the intelligent (and quasi-intelligent) spin
states introduced by Aragone et al. We use these states to construct
families of coherent wave packets on the sphere and we sketch the time
evolution of these wave packets for a rigid body molecule. 
\end{abstract}
\submitted 

\section{Introduction}
  The eigenstates of the square of the angular momentum operator $L^2$
which are also eigenstates of $\Lp$,
where $\eta$ is a real parameter,
have been called the {\em intelligent spin states} \cite{aragone} %[1] 
and have been the subject of intensive analytical studies 
\cite{kolodz,rashid}. %[2],[3]. 
However, at least to our
knowledge, one does not find in the literature any discussion of their
geometrical interpretation and of their utility in the construction of
angular momentum coherent states. 
In completing a recent work 
\cite{rozmej} %[4] 
on the time evolution of coherent states built from a subclass 
of those states we
came to conclusion that such a discussion had still to be presented. 

     In this short article we will first show in section %2 
\ref{ssots}     that the
parameter $\eta$ enables to define squeezed states on the sphere i.e. 
states for which the uncertainties 
$\Delta L_x^2$ and $\Delta L_y^2$ can be varied at will.
It is however for real values of $\eta$ that the product of the
uncertainties has a minimum value. In section %3 
\ref{cotss} we will show that the
states can be classified into three categories, a result already found in
ref.~\cite{rashid}, %[3], 
and that the well known Radcliffe's states belong to one of 
these categories. In section %4 
\ref{soafoewp} we will study the angular localization 
and the partial wave expansion of a particular class generated from 
an exponential coherent state on the sphere. 
This state was introduced by us elsewhere~\cite{rozmej}. %[4].
Finally in section %5 
\ref{frftcoarr} we will study the time evolution of this
particular set of states in the same line as in \cite{rozmej} %[4] 
i.e.  assuming that the wave packets evolve with the hamiltonian of 
a rigid body with moment of inertia $J$ i.e. 
\begin{equation}\label{Ham} %(1)
            H = \frac{\hbar^2}{2J}\,L^2  \: .  
\end{equation}
   In particular we will show that the scenario of fractional revivals
found in \cite{averbukh} %[8] 
is well exhibited by the family of wave packets studied in
section \ref{soafoewp}.

\section{Squeezed states on the sphere}\label{ssots}
%{\bf a)-} 
   Let us enumerate a few general properties of the states  
$|w,\eta\rangle$  which
are eigenstates of $\Lp$ with a complex value of $\eta$. In this first
part there is no need to assume that they are eigenstates of $L^2$.
These states are the normalized states which obey the equation:
\begin{equation}\label{l1} %(1)
(\Lp)|w,\eta\rangle = w|w,\eta\rangle   \: .
\end{equation}
				     
 It is a simple exercise, already discussed in references \cite{jackiw} 
 and \cite{levy}, % [5] and [6], 
 to prove equations (\ref{l2}-\ref{l7}). First of all 
\begin{equation}\label{l2} %(2) 
|\eta|^2 = \frac{\Delta L_x^2}{\Delta L_y^2}   \: ,
\end{equation}				     
i.e. $|\eta|$ can be called the squeezing parameter (with the definitions 
$\Delta L_i^2=\langle L_i^2\rangle -\langle L_i\rangle^2, i=x,y$). The
phase $\alpha$ of $\eta$ determines the ratio of the average value of the
anticommutator of $L_x$ with $L_y$ to the average of their 
commutator since
\begin{equation}\label{l3} %(3)
\tan\alpha = \frac{\langle\{ L_x,L_y \}\rangle - \langle L_x\rangle
\langle L_y\rangle}{\langle L_z\rangle}   \: . 
\end{equation}				     
Finally the product of the uncertainties is given by: 
\begin{equation}\label{l4} %(4)
\Delta L_x^2 \Delta L_y^2 = \frac{1}{4}[\langle L_z\rangle^2 +
|\langle\{ L_x,L_y \}\rangle - \langle L_x\rangle\langle L_y\rangle|^2]
= \frac{1}{4} \frac{\langle L_z\rangle^2}{\cos^2\alpha}   \: .
\end{equation}				     
The average values of $L_x$ and $L_y$ are fixed both by the 
parameter $\eta$  and the eigenvalue $w$ by:
\begin{equation}\label{l5} %(5)
\langle L_x\rangle = \frac{\eta w^*+w\eta^*}{\eta+\eta^*}, \qquad
\langle L_y\rangle = \frac {1}{i}\frac{w-w^*}{\eta+\eta^*}   \: .
\end{equation}				     
 The eigenstates corresponding to a real parameter $\eta$ satisfy the
important minimum uncertainty relation from which the work 
\cite{aragone} was initiated
\begin{equation}\label{l7} %(7)
\Delta L_x^2 \Delta L_y^2 = \frac{1}{4}\langle L_z\rangle^2   \: .
\end{equation}				     
 
\section{Classification of the squeezed states}\label{cotss}
%{\bf b)-} 
The eigenvalue $w$ can be obtained very simply in the basis of
eigenstates of $L^2$ with eigenvalue $l(l+1)$ if one uses the 
observation \cite{rashid} %[3]  
that, within a constant factor $\sqrt{1-\eta^2}$, 
the operator $\Lp$ is one of the three generators of an SU(2) algebra. 
The set of operators satisfying this algebra is defined as
\begin{equation}\label{l8} %(8)-(9)
{\cal L}_3 = \frac{\Lp}{\sqe}, \qquad
%\end{equation}				     
%\begin{equation}\label{l9} %(8)-(9)
{\cal L}_{\pm} = \pm \left(\frac{\eta L_x +iL_y}{\sqe}\right) - L_z  \: .
\end{equation}				     
Therefore there are $2l+1$ solutions to equation (\ref{l1});
instead of $w$ one uses simply
the eigenvalue of ${\cal L}_3$ with $k=-l,\ldots ,+l$ and the formula
\begin{equation}\label{l10} %(10)
w = k\sqe   \: .
\end{equation}				     
 Let us denote by $|l,k,\eta\rangle$ the eigenstates solutions of 
 equation (\ref{l1}) in the basis where $L^2$ is diagonal.
(Note that $k$ is the eigenvalue of ${\cal L}_3$ and not of $L_z$).
  Using (\ref{l10}) one obtains two expressions for the average of $L_x$
and $L_y$ in the particular case of a real value of $\eta$: 
\begin{equation}\label{l11} %(11) 
\langle L_x\rangle = k\sqe , \qquad \langle L_y\rangle=0 , \qquad  
\mbox{if} \qquad |\eta|<1
\end{equation}				     
\begin{equation}\label{l12} %(12) 
\langle L_x\rangle = 0 , \qquad \langle L_y\rangle=k\sqrt{\eta^2-1} , \qquad  
\mbox{if} \qquad |\eta|>1  \: .
\end{equation}				     
The case with $|\eta|=1$ is obviously singular but there is a unique well
known solution for which $w=0$. One has then  
\begin{equation}\label{l13} %(13)
(L_x\pm iL_y) |l,k=\pm l,\eta=\pm 1\rangle = 0  
\end{equation}				     
 and $|l,\pm l,\eta=\pm 1 \rangle$ coincide with the
eigenstates of $L_z$ with eigenvalues $\pm l$. 
  If $\eta$ takes a complex value 
$\langle L_x \rangle$ and $\langle L_y \rangle$ are both nonzero but are 
both proportional to $k$.
  The intelligent spin states are the solutions of 
equation (\ref{l1}) with average
values given by (\ref{l11}) or (\ref{l12}) 
and which satisfy moreover equation (\ref{l7}). The
{\em generalized intelligent states} or the 
{\em quasi-intelligent spin states}
which were respectively defined in \cite{aragone} and \cite{rashid}
% [1] and[3] 
are the solutions of equation (\ref{l1})
with a complex value of $\eta$ for which both averages are nonzero and to
which equation (\ref{l4}) must be applied with a nonzero value of $\alpha$.
In both cases it is sufficient to consider the interval $|\eta|<1$
and $\alpha \in [0,\pi/2]$.

   Let us now classify those states according to $k$ as follows:
\begin{itemize}    
\item[~~i)] The states with $k=0$ for which $\langle L_x \rangle$  and 
     $\langle L_y \rangle$ are zero and only $\langle L_z \rangle$
is nonzero.\\ 
\item[ ii)] The states with $k=\pm l$ which are very particular 
     as we will show below. \\
\item[iii)]  The states with intermediate values of $k$.
\end{itemize}
    
  In order to justify this classification we have to express 
  $\langle L_z \rangle$ as
\begin{equation}\label{l14} %(14)  
\langle L_z\rangle = \eta \langle{\cal L}_3\rangle +i\sqe 
\langle L_y\rangle - \langle{\cal L}_+\rangle   \: .
\end{equation}				     
Except for the case $k=\pm l$ the average of the
operator $\Lp$ defined by equation (\ref{l8}) is nonzero. 
This average can be calculated from the works of
\cite{aragone} or \cite{rashid}. %[1] or [3]. 
It is generally the ratio between two
polynomials of the parameter $\eta$, the degree of which increases with
$l$. Therefore the angle between the vector $\langle \vec{L} \rangle$
and the $Oz$ axis is not a simple function of $\eta$ for general values 
of $k$ and $l$ and there is no
simple geometrical meaning of this angle. For $k=\pm l$ such an
interpretation can indeed be found. Let us define $\eta$ in terms of two
angles $\theta_0$ and $\phi_0$ by
\begin{equation}\label{l15} %(15)
\eta = \frac{\tan\phi_0+i\cos\theta_0}{\cos\theta_0\tan\phi_0+i}  \: .
\end{equation}				     
With this value of $\eta$ the equation 
\begin{equation}\label{l16} %(16)
(\Lp) |l,l,\eta \rangle = l\sqe |l,l,\eta \rangle 
\end{equation}				     
 can be written simply as       
\begin{equation}\label{l17} %(17)
(\vec{L}\cdot\vec{l}) |l,l,\eta \rangle = l |l,l,\eta \rangle   \: ,
\end{equation}				     
 where the vector $\vec{l}$ is the unit vector in the direction of 
 $\langle \vec{L} \rangle$  which is such that
\begin{equation}\label{l18} %(18) 
\langle L_z\rangle = l\sin\theta_0\cos\phi_0, \quad
\langle L_y\rangle = l\sin\theta_0\sin\phi_0, \quad
\langle L_z\rangle = l\cos\theta_0   \: .
\end{equation}				     
  Equation (\ref{l18}) expresses the fact that $|l,l,\eta\rangle$ 
  are simply spherical
harmonics $Y_l^l$ in a system of axis where $Oz$ is along the vector $l$.
These states are called generally Radcliffe's states \cite{radcliffe}. 
% [7]. 
Let $\vec{u},\vec{v},\vec{l}$ be three unit
vectors forming an orthogonal direct system. One has also
\begin{equation}\label{l19} %(19) 
 (\vec{L}\cdot (\vec{u}+i\vec{v})|l,l,\eta \rangle = 0    \: .    
\end{equation}				     
 In this rotated system of coordinates the parameter $\eta$ 
 is equal to one!\footnote{One also finds other cases where the states 
coincide with spherical harmonics with an axis of quantization 
different from $Oz$: it is the $Ox$ axis
for $\eta=0$ and for $\eta=i|\eta|$ this axis has $\theta=\pi/2$ and
$\sin\phi=-|\eta|/\sqrt{1-\eta^2}$.} 
 
    For $k \neq \pm l$ the transformation which enables to build up
the intelligent or quasi-intelligent spin states from the usual spherical
harmonics is not a rotation. This remark was already formulated 
long ago in \cite{rashid}  %[3] 
at the same time as the previous remark on the 
equivalence of Radcliffe's states with the states with $k=\pm l$.  
 
\section{Study of a family of exponential wave packets}\label{soafoewp}
% {\bf c)-}
  Let us now discuss the properties of wave packets built from
intelligent or quasi-intelligent spin states with the same value of $w$, 
i.e. of $k$, and containing many different values of $l$. 
The consideration of such
admixtures enables indeed to modify at our will the angular spread keeping
either (\ref{l4})  or (\ref{l7}) 
and with the allowance, provided by equation (\ref{l2}),
that $\eta$ is an adjustable squeezing parameter.
These WP can be expressed either in the
basis $|l,k,\eta\rangle$ with coefficients $c_l(k,\eta)$ as
\begin{equation}\label{l20} %(20)
|\Psi_{\eta k}\rangle = \sum_l \, c_l(k,\eta) |l,k,\eta\rangle
\end{equation}				     
 or in the basis of ordinary spherical harmonics with coefficients
$b_{lm}(k,\eta)$ as
\begin{equation}\label{l21} %(21) 
\Psi_{\eta k}(\theta,\phi) = \sum_{lm}\,b_{lm}(k,\eta) \,
Y_m^l(\theta,\phi)   \: .  
\end{equation}				     
We will discuss from now on the properties of families of states
generated from a {\em parent} state investigated by us elsewhere 
\cite{rozmej}  and %[4] and
called the {\em exponential coherent state}. For a real value of $\eta$ 
this state depends on a single parameter $N$ and a function of 
$\theta$ and $\phi$ called $v$, 
its expression is: 
\begin{equation}\label{l22} %(22) 
\Psi_{\eta 0}(\theta,\phi) = \sqrt{\frac{N}{2\pi\sinh 2N}} e^{Nv}
= \sqrt{\frac{N}{2\pi\sinh 2N}} e^{N\sin\theta(\cos\phi+i\eta\sin\phi)}
\: . 
\end{equation}				     
 The probability density which is associated depends only on $N$. 
It is maximum within a solid angle symmetric
around the axis $Ox$ and the width of this solid angle is of the order
$4\pi/(4N+1)$. Also the average  $\langle L_z \rangle$
obeys the following formula
\begin{equation}\label{l23} %(23)
\langle L_z\rangle = \eta(N\coth (2N)-\frac{1}{2}) \mylims
\eta(N-\frac{1}{2})\: .
\end{equation}				     
 We have shown in \cite{rozmej} %[4] 
how to obtain this WP from a three dimensional harmonic oscillator 
coherent state. 
As said in \cite{rozmej} %ref[4]  
all the WP have $k=0$. 
  Our purpose is now to construct for real or complex $\eta$ 
families of WP having increasing values of $k$
using (\ref{l22}) for the parent state. We would
like to study the angular spread as a function of $k$ and so to say the
coherence properties of the following family: 
\begin{equation}\label{l24} %(24)
\Psi_{\eta 0},\quad {\cal L}_+ \Psi_{\eta 0},\quad 
{\cal L}_+^2 \Psi_{\eta 0}, \ldots ,
 {\cal L}_+^k \Psi_{\eta 0}   \: .
\end{equation}				     
  These states are all solutions of equation (\ref{l1}) 
  with respectively eigenvalues
 $0, \sqrt{1-\eta^2}, 2\sqrt{1-\eta^2}, \ldots ,k \sqrt{1-\eta^2},\ldots $
 If $\eta$ is real they all satisfy equation (\ref{l7}) 
 while equation (\ref{l4}) is satisfied 
 if $\eta$ is complex. 
Applying ${\cal L}_+$ onto the argument $v$ of (\ref{l22}) one obtains a
 function $v_+$ defined by
\begin{equation}\label{l25} %(25) 
\fl
v_+ = {\cal L}_+ v = \frac{1}{2}(\cos\theta-\eta)\sqrt{\frac{1+\eta}{1-\eta}}
- \frac{1}{2}(\cos\theta+\eta)\sqrt{\frac{1-\eta}{1+\eta}}
+ (\eta\cos\phi+i\sin\phi)\sin\theta   \: .
\end{equation}				     
But a second application leads to the
following property:                                                     
\begin{equation}\label{l26} %(26)
 {\cal L}_+ v_+ = {\cal L}_+^2 v = 0 \: .
\end{equation}				     
 The set defined by (\ref{l24}) can be expressed then simply in terms of 
 $\Psi_{\eta,0}$
and of powers of $v_+$ as
\begin{equation}\label{l27} %(27)  
\Psi_{\eta 0},\quad  v_+ \Psi_{\eta 0},\quad v_+^2 \Psi_{\eta 0}, \ldots ,
 v_+^k \Psi_{\eta 0} \: .
\end{equation}				     
 In this manner one sees that the
gaussian will be dominant for any $k$ in such a way that the WP will be
highly concentrated on the sphere even for high $k$. The action of the
operator ${\cal L}_+^k$ is however very different if one analyses the 
decomposition of WP into
partial waves according to the expansion (\ref{l21}).
Indeed it suppress all the
partial waves with $l<k$, it moves the distribution towards the values with
$m>l$ and also, by the change in the normalization, 
it increases the weight of the higher partial waves. 
However if one uses not the basis of the usual spherical harmonics 
(\ref{l21}) but the basis of intelligent spin states one can
draw benefit from the fact that we have in this basis the equation
\begin{equation}\label{l28} %(28)
 {\cal L}_+ |l,k,\eta\rangle = \sqrt{l(l+1)-k(k+1)} |l,k+1,\eta\rangle 
\end{equation}				     
 which shows that the relative phases of
the states present in (\ref{l20}) are not affected.
Therefore the states will keep
their coherence property as a function of $k$.
 
    To be complete we give below the expression of $b_{lm}(0,\eta)$ 
    from which a
recurrence can be built easily to find the coefficients for $k>0$.
One has found in \cite{rozmej} that for real values of $\eta$
\begin{equation}\label{l29} %(29)
\fl
b_{lm} = \sqrt{\frac{2N}{\sinh (2N)}} \, \sum_{l_1l_2} \,(-1)^{l_2} \,
\frac{(N(1+\eta))^{l_1} (N(1-\eta))^{l_2}}{\sqrt{(2l_1)!(2l_2)!}} \,
\frac{\langle l_1 l_2 00|l0\rangle 
\langle l_1 l_2 l_1 -l_2|lm\rangle}{\sqrt{(2l+1)}}\; .
\end{equation}				     
For complex $\eta$ the square root in front must be changed.

 The distribution of the coefficients $b_{lm}$ is presented 
in figure 1 for $\eta=0.5, N=0$ and $k=0,10$ and 20. 
A very large shift of the distribution of
the $b$'s towards the higher $l$ and $m>l$ is clearly observed when $k$ 
increases as well as an increase in the spread of the values of $m$. 
  
   Sections of the WP at their maximum are presented in figure 2  
which shows the angular shape of the WP. 
Apart from a small movement towards smaller
values of $\theta$ when $k$ increases the WP is almost as strongly 
localized for $k=20$ as it was for $k=0$ for $\eta=0.5$ and
$\alpha=0$, as well as for $\alpha\neq 0$.

    Despite the strong angular concentration of the WP for $k=20$ the
calculation of $\langle L_z \rangle$
provides very large values for the product of the
uncertainties. This average value is shown in figure 3 for $\eta=0.5$ 
in the same time as $\langle L^2 \rangle$. 
  
\section{Fractional revivals for the case of a rigid
rotation}\label{frftcoarr}
  The time evolution of the wave packets described above can now be
studied assuming the hamiltonian (\ref{Ham}).
This assumption was applied in \cite{rozmej} %[4]
for the state (\ref{l22}), i.e the state representing a
rigid heteronuclear molecule and we have made an
extensive study of its time evolution. 
The fractional waves
that one obtains as time proceeds are obtained from the initial WP by
multiplying the $b_{lm}$ by 
$\exp (-iI\omega_0 t_s)$  (see \cite{rozmej}) where $\omega_0$ is the
frequency of periodicity of the rigid rotor and where $t_s$ is a 
fractional time. 
It is this change of $t_s$ which allows to obtain a rich variety of
fractional waves on the sphere for $k=0$.
Depending on the value of $\eta$ 
(only real values of $\eta$ were considered in \cite{rozmej}) %[4]) 
the WP exhibits a rich scenario
of fractional revivals in the line described in \cite{averbukh}, %[8], 
the origin of which being traced in the quantum mechanical spreading. 
The parameter $\eta$ allows
to control the relative spread of each of the angular variables. For
fractional time $(m/n)\,T_{rev}$, where $T_{rev}$ is a common revival time, 
the WP is subdivided into a certain number of WP ($n$ if $n$ is odd, 
$n/2$ in the even case) 
the shape of which depend strongly on the value of $\eta$. For the case
$\eta=\pm 1$ the fractional WP are clones of the initial WP. 
For different
values their shape change (we have called these WP {\em mutants}). 
If $N$ is large enough the fractional WP are located around 
different directions on the sphere and do not interfere much spatially 
for low enough values of $n$ ($n<8$ for $N=20$). 
We interpret this properties as a manifestation of a robust
virtue of coherence of the WP. 
The revivals of various WP having $k=0,5,10$ and 20 are presented for 
$m/n=1/10$, such a time has been chosen as a typical exemple. 
There we should observe 5 wave packets according to 
 \cite{averbukh}. %[8]. 
These WP are always very well separated from each other,
again there is no significant difference between the WPs constructed
from intelligent spin states and those built from 
quasi-intelligent ones.
  
\section{Conclusions}\label{con}
  The conclusion of this article is that there exists very numerous
possibilities of construction of angular coherent states using the
properties of the intelligent spin states. 
For a system with an hamiltonian quadratic in $I$ [$I(I+1)$ spectrum] 
these WP spread on the sphere but there
is a well identified mechanism of fractional revivals that produce a set
of well concentrated {\em mutants}. For rotators which are not 
quadratic the scenario is valid during a limited time. 
This is the case of nuclei for which we are making a parallel study
\cite{rozmejapp}. %[9]  

\ackn
One of us (P.R.) kindly acknowledge support of Polish Committee for Scientific 
Research (KBN) under the grant 2 P03B 143 14.

\end{document}